\documentclass[12pt,a4paper]{article}

\usepackage{amsfonts}
\usepackage{graphics}

\newcommand{\be}{\begin{equation}}
\newcommand{\ee}{\end{equation}}
\newcommand{\eel}[1]{\label{#1}\end{equation}}
\newcommand{\bea}{\begin{eqnarray}}
\newcommand{\eea}{\end{eqnarray}}
\newcommand{\eeal}[1]{\label{#1}\end{eqnarray}}
\newcommand{\baq}{\begin{equation}\begin{array}{rcl}}
\newcommand{\eaq}{\end{array}\end{equation}}
\newcommand{\eaql}[1]{\end{array}\label{#1}\end{equation}}
\newcommand{\beac}{\begin{equation}\begin{array}{rcl}}
\newcommand{\eeacn}[1]{\end{array}\label{#1}\end{equation}}
\newcommand{\ba}{\begin{array}}
\newcommand{\ea}{\end{array}}

\newcommand{\equ}[1]{(\ref{#1})}
\renewcommand{\a}{\alpha}

\newcommand{\al}{{\alpha^{'}}}
\newcommand{\beq}{\begin{eqnarray}}
\newcommand{\eeq}{\end{eqnarray}}

\newcommand{\journal}[4]{{\rm #1~}{#2}\,(#3)\,#4}

\newcommand{\np}{\journal {Nucl. Phys.}}

\newcommand{\gym}{g_{YM}}

\begin{document}
\newcommand{\preprint}[1]{\begin{table}[t]  %%
           \begin{flushright}               %%
           \begin{large}{#1}\end{large}     %%
           \end{flushright}                 %%
           \end{table}}                     %%
\preprint{hep-th/9806158\\TAUP-2505-98}

\begin{center}
\LARGE{Baryons from Supergravity}

\vspace{10mm}

\normalsize{A. Brandhuber, N. Itzhaki,   
J. Sonnenschein and S. Yankielowicz}

\vspace{10mm}

{\em  Tel Aviv University, Ramat Aviv, 69978, Israel\\
andreasb, sanny, cobi, shimonya@post.tau.ac.il}
\end{center}

\vspace{10mm}

\begin{abstract}
We study the construction of baryons via supergravity along the line 
suggested recently by Witten and by Gross and Ooguri.
We calculate the energy of the baryon as a function of its size.
As expected the energy is linear with $N$. 
For the non-supersymmetric theories (in three and four dimensions) we
find  a linear relation which is an indication of confinement.  
For the ${\cal N} = 4$ theory  we 
obtain  the result ($E L= - \mbox{const.}$) which is compatible with
conformal invariance.
Surprisingly, our calculation suggests that there is a bound state of
$k$ quarks if $N\geq k\geq 5N/8$. We study the ${\cal N} = 4$ theory
also at finite temperature and find the zero temperature behavior for
small size of the baryon, and screening behavior
for  baryon, whose size is  large
compared to the thermal wavelength.

\end{abstract}

\newpage

\section{Introduction}
Recently, \cite{juan} it was conjectured that
four-dimensional $\mathcal{N} = 4$ supersymmetric Yang-Mills theory with
gauge group $SU(N)$ is dual to type IIB string theory on the
background $AdS_5 \times S^5$ (where AdS is five-dimensional
anti-de-Sitter space). In this correspondence the string coupling
$g_s$ is equal to the gauge coupling $g_{YM}^2$ and $(g_{YM}^2
N)^{1/4} \equiv (g_{eff}^2)^{1/4}$ is proportional to the radius of the
AdS space and the five-sphere (in string units). There are $N$ units of
five-form flux on the $S^5$ in string theory. In the limit of large
$N$ and large $g_{eff}^2$ the string theory is reliably approximated
by supergravity and one expects to be able to extract gauge theory
correlations functions, the set of chiral operators and the mass-spectrum of
the strongly coupled gauge theory using classical calculations in
supergravity. 

A precise relationship between the supergravity effective action and
gauge theory correlators \cite{gkp, witten1} and a match of chiral 
primary operators in
the conformal theory with Kaluza-Klein states of the compactified
supergravity \cite{witten1} has been found. Likewise, there exist precise
recipes for computing 
%correlation functions of 
the Wilson loop operators
\cite{sjr,juan1}. 
This allows us  to study the qualitative behavior of gauge
theories (confinement, screening, ...)  at zero or
finite temperature and also of non-supersymmetric theories
\cite{witten2,bisy1,bisy2}. The
relevant configuration of external quark and anti-quark can be thought
of as a mesonic vertex operators. In the supergravity description they are
constructed as an open fundamental string connecting to separated
points on the boundary of AdS space where the string endpoints
correspond to the external quark and anti-quark, respectively.

More recently the construction of baryons was discussed in the
supergravity framework \cite{witten,og}. 
The precise meaning of a baryon in
this context is a finite energy configuration of $N$ external
quarks. The $\mathcal{N} = 4$ SYM theory does not contain dynamical quarks
in the fundamental representation which are necessary to construct
baryonic particles. (An interesting counter example to this is the
``Pfaffian'' particle which is constructed out of adjoint fields in
$\mathcal{N} = 4$ with $SO(2 N)$ gauge group \cite{witten}).

In the construction of a baryon vertex in string theory one faces a
puzzle. If we think of the $N$ quarks as endpoints on the AdS boundary
of $N$ fundamental strings with equal orientation it seems a priori
inconsistent to let the other ends of the strings terminate on one
point in the interior of AdS. Nevertheless, it was shown in \cite{witten}
that this is possible (see \cite{og} for a different argument involving
the Chern-Simons term of the compactified supergravity). 
The baryon vertex turns out to be a
D5 brane wrapped on the $S^5$. In the type IIB
string theory there is a self-dual field strength $G_5$ and, as
mentioned earlier, the compactification on $AdS_5 \times S^5$ has $N$
units of flux on the five-sphere: $\int_{S^5} \frac{G_5}{2 \pi} = N$. On the
D5 brane world volume there is a $U(1)$ gauge field $A$ which couples to the
five-form field strength through the term $\int_{R\times S^5} A \wedge
\frac{G_5}{2 \pi}$. Because of this coupling $G_5$ contributes $N$
units of $U(1)$ charge. Each string endpoint adds $-1$ unit of charge
Since in a compact space the total charge has to vanish, precisely
$N$ strings have to end on the D5 brane. In the $SU(N)$ gauge theory
the gauge invariant combination of $N$ quarks is completely
antisymmetric and, indeed, the strings between the boundary (or a D3
brane) and the D5 brane are fermionic strings \cite{witten} because the
strings have mixed DN boundary conditions in eight space directions.

In the supergravity description of non-conformal theories \cite{imsy} we 
can use similar arguments. The starting point is a set of $N$
D$p$-branes which give rise to $N$ units of flux of a $p+2$-form field
strength $G_{p+2}$ of the type II string theory: 
$\frac{1}{2 \pi} \int_{S^{8-p}} \star G_{p+2} = N$. 
The baryonic vertex is represented by a D$(8-p)$-brane wrapped on an
$S^{8-p}$ with $U$ dependent radius. The $U(1)$ gauge field $A$ on the
D$(8-p)$-brane couples to $G_{p+2}$ through the term $\frac{1}{2 \pi}
\int_{R\times S^{8-p}} A \wedge \star G_{p+2}$ and, therefore, leads
to $N$ units of $U(1)$ charge which are canceled by $-N$ units of
charge from $N$ fundamental strings ending on the D$(8-p)$-brane.

In the present letter we want to study baryonic vertices in
detail and calculate the energies of such configurations in string
theory. The energy
has two main contributions, the tension of the strings and the energy
of the D5 brane (we will not discuss corrections due to
interactions between the strings). Both contributions are proportional
to $N$ but come with opposite signs. Therefore, stable configurations
exist and we calculate the energy as a function of the characteristic
size $L$ of the baryon. For a static configuration we have
to demand that the net force on the vertex vanishes. Using this constraint
we determine the angle at which the strings end on the vertex. We
study baryon  in 
four dimensional theories with maximal supersymmetry at zero and
finite temperature and non-supersymmetric theories in three and four
dimensions and compare the results to Wilson loop calculations.
Furthermore, our calculation shows that there is a bound state of $k$
quarks if $k$ satisfies $N\geq k\geq 5N/8$.
The expression for the baryon energy in maximally supersymmetric 
theories of other dimensions is also written down.

%\section{ Baryons of ${\cal N}=4$ SYM in four dimensions}

\section{ Baryons of ${\cal N}=4$ SYM in four dimensions}

%What we would like to do is to elaborate on the supergravity
%construction of the baryons.
%In particular, we are after the relation between 
%the location of the vertex at the bulk and the size and energy
%of the baryon.

Consider the baryon configuration suggested in \cite{witten}.
There are two contributions to the action of the system.
The first contribution comes from the string stretched between the boundary  of the AdS$_5$ space
and the
D5-brane wrapped on the $S_5$.
 The second contribution comes from the D5-branes itself.
As was noted in \cite{witten} they are of the same order.
 Hence, we  should
consider both of them.
Let us start with the D5-brane.
Since we are considering a static D5-brane  wrapped on $S^5$ its
action is 
\be
S_{D5}=\frac{1}{(2\pi)^5 \al ^3 e^{\phi}}\int dx^6 \sqrt
{h}=\frac{T NU_0 }{8\pi},
\ee 
where $U_0$ is the location of the baryon vertex in the bulk, $T$ is
the time period which we take to infinity and 
 $h$ is the induced metric on the fivebrane.

The configuration which we consider ( see Fig. 1)  is such that the strings end on a
surface with radius $L$ in a symmetric way which ensures that the net
force on the vertex along $x_i$ vanishes (where $x_i$ are the
direction along the boundary where the field theory is living).
 Hence the configuration is
stable in the $x_i$ directions.
Of course to  stabilize the system  along the $U$ direction the
symmetry
argument 
is not enough, one has to consider the strings action as well.

\begin{figure}
\begin{center}
 \resizebox{8cm}{!}{
   \includegraphics{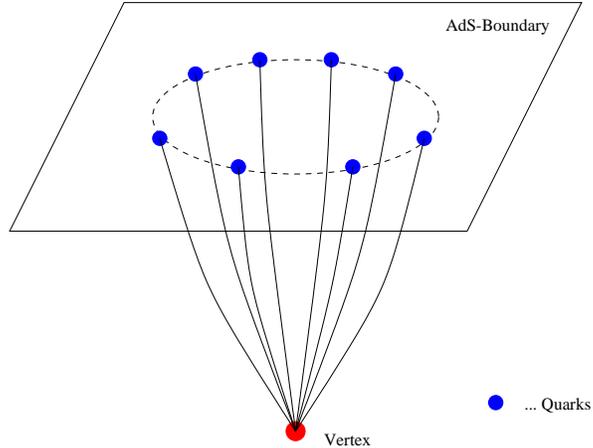}
   }
\end{center}
\caption{The Baryon Vertex}
\end{figure}

Following \cite{juan1}, we work with the Nambu Goto action in the
gauge $x=\sigma$ and $t=\tau$ which gives
\be
S_{1F}=\frac{T}{2\pi}\int dx \sqrt{U_x^2+U^4/R^4},
\ee
where $U_x=\frac{\partial U}{\partial x}$ and $R^4=4\pi g_s N$.
The total action is
\be\label{action}
S_{total}=S_{D5}+N S_{1F}.
\ee
The variation of (\ref{action}) under $U\rightarrow U +\delta U$
contains a volume term as well as a surface term.
The volume term leads to the Euler-Lagrange equation whose solution
satisfies \cite{juan1} 
\be\label{1}
\frac{U^4}{\sqrt{U_x^2+U^4/R^4}}= \mbox{const.}
\ee
because the action does not depend explicitly on $x$.

The surface term yields
\be\label{2}
\delta U\frac{TN  }{8\pi} 
=\delta U\frac{TN(U_x)_0}{2\pi\sqrt{(U_x)_0^2+U_0^4/R^4}},
\ee
where $(U_x)_0= \frac{\partial U}{\partial x}|_{U_0}$
and $\delta U$ is the variation of $U$ at $x=0$ where the string
hits the baryon vertex. 
This condition  is simply the no-force condition in the curved space-time.
%Denoting by $U_0$ the location of the D5-brane and
 Using (\ref{1}) and
(\ref{2}) one finds that
\be
\frac{U^4}{\sqrt{U_x^2+U^4/R^4}}= \sqrt{\frac{15}{16}} U_0^2R^2.
\ee
This   implies  the following relation between $U_0$ and the
radius of the  baryon $L$ 
\be
L = \frac{R^2}{U_0}  \int_1^{\infty}
\frac{dy}{y^2\sqrt{(\beta^2 y^4 - 1)}} ~
\eel{length}
where $\beta=\sqrt{16/15}$. 
The energy of a single string is given by 
\be\label{energyst}
E = \frac{1}{2\pi}U_0\left(  \int_1^\infty dy
\frac{\beta y^2}{\sqrt{\beta^2 y^4 - 1}} - 1 \right)
- \frac{U_0}{2\pi} 
\ee
Where we subtract the energy of the configuration with the D5-brane 
located at $U=0$.
Since  $g_{xx}$ vanishes at $U=0$  any radial string  which reaches
this point ends on the D5-brane.
As a result the energy of the fermionic strings, which we subtract
equals
  the energy of free
quarks\footnote{By free quarks we mean string which are stretched from
  $U=\infty$ to $U=0$}.
Note that since $g_{tt}(U=0)=0$ the contribution  of the D5-brane located at
$U=0$ to the energy  vanishes.
%Therefore, we subtract  only  the energy of $N$ free quarks. 
% located at $U=0$ is located at the
%The subtraction we have introduced seems to be 
%the same as the one used for  the 
%quark anti-quark case\cite{juan1}, namely, subtracting the masses of the 
% individual quarks. In the AdS picture one may conclude that
%for the baryon configuration  we subtract  the energy associated with
%$N$ straight not connected  
%strings stretched  from $U=0$ to $U\rightarrow \infty$.
%In fact the reference system is that of $N$ strings connected to a baryon 
%vertex located at the origin similar to the one drawn in Fig 3.b.
% Note that  the invariant energy of the $D_5$ brane
%at the origin vanishes the    

Inserting the relation \equ{length} into \equ{energyst} 
%and \equ{energy5}
one finds  that the  energy of each string is 
\be
E= -\a_{st} {\sqrt{2\gym^2 N  }\over L},~~\mbox{where}~~\a_{st}=~ 
\frac{1}{4} \sqrt{\frac{5}{6 \pi}} {}_2F_{1}[ \frac{1}{2}
, \frac{3}{4}, \frac{7}{4}; \frac{15}{16}] \times
{}_2F_{1}[\frac{1}{2}, - \frac{1}{4}, \frac{3}{4}; 
\frac{15}{16}]\simeq 0.036 
\eel{energytots}
The total energy of the baryon configuration is therefore
\be
E= -\a_B N {\sqrt{2\gym^2 N  }\over L},~~\mbox{where}~~\a_B=...\simeq 0.007 
\eel{energytotB}

Since the force $F=\frac{dE}{dL}$ is positive  the 
baryon configuration is stable.
 Moreover, as expected from 
the  field theory large N analysis, the energy  is proportional to $N$ 
times that of the quark anti-quark system.    
Recall \cite{juan1}  that while the fact    
that the energy is proportional to $1/L$  is dictated by 
 conformal invariance, the  
 dependence on $R^2$ is a non-trivial prediction of the  AdS formulation
concerning the strong coupling behavior of the gauge theory.

% however, is not determined by the conformal
%invariance of the theory thus it is worth while to note that the $R^2$
%dependence is the same as in the Wilson line \cite{juan1} thought
%the numerical coefficient is different ****.

{\it ${\cal N}=4$ at finite temperature}

In \cite{bisy1,rty}
 the Wilson loop at finite temperature was considered.
The temperature introduces a scale into the conformal theory,
which distinguish between the behavior in two regions.
So there are two regions.
At distances smaller then the wavelength associated with the
temperature the behavior is essentially the conformal one ($E\sim -1/L$)
with corrections.
However, at large distances the charges are screened by the the
effects of the temperature.
>From the supergravity point of view what is happening is the
following. 
In the $T=0$ case to increase $L$ one has to decrease
$U_0$ (the point where the slope , $U_x$, is zero).
In the presence of a temperature one cannot decrease $U_0$ below the
horizon, associated with the temperature, and hence it seems that we
have a maximal distance of separation between the quark and the
anti-quark. 
However, before we reach that point the energy becomes positive (after
the subtraction of the free quarks energy) and hence the quark
anti-quark system becomes free.
\begin{figure}
\begin{center}
 \resizebox{8cm}{!}{
   \includegraphics{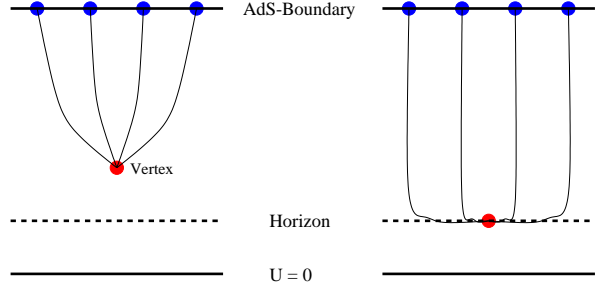}
   }
\end{center}
\caption{The two ways to obtains $N$ fermionic quarks at the boundary.
At small distances (compared to the wavelength of the temperature)
the lowest energy configuration is the one with
the D5-brane above the horizon. 
In this configuration there is a
potential between the quarks and the vertex.
At large distances the lowest energy configuration  is the one with
the D5-brane located at the horizon.
Since $g_{tt}$ vanishes at the horizon the contribution to the energy
from the horizontal parts of the strings (as well as from the
D5-brane) vanishes. Hence, the energy
does not depend on the size of the baryon.
In other words the energy of this configuration is simply $N$ times 
the energy of a string
stretched between $U=\infty$ and the horizon which explains the
subtraction  made at eq.(\ref{ui}).}

\end{figure}
The situation with the baryons is rather similar although there are
some technical differences.
The surface term now yields,
\be
\frac{U_0'}{\sqrt{(U_0')^2 + (U_0^4 - U_T^4)/R^4}} = \frac{1}{4}
\left(\frac{1 + U_T^4/U_0^4}{\sqrt{1-U_T^4/U_0^4}} \right) ~.
\ee
We see that the minimal value of $U_0$ is not the location of the
 horizon $U_t$ but $\gamma U_0$ where $\gamma >1$.
The integral for the size $L$  baryon takes the
following form
\be
L = \frac{R^2}{U_0} \int_1^\infty dy \sqrt{ \frac{15 - 18 \rho^4 -
    \rho^8}{(y^4 - \rho^4)(16 y^4 - 15 + 2 \rho^4 + \rho^8)}  }
\ee
where $\rho = U_T/U_0$.
Since the minimal $U_0$ is larger then in the Wilson loop case the
maximal $L$ is smaller than in the Wilson loop case.
Thus one might worry that we reach $L_{max}$ before we reach the 
 positive energy condition.
However, now the energy of the system contains also a positive term
coming from the D5-brane\footnote{For an explanation on the
  subtraction see figure 2.} 
\be\label{ui}
E = \frac{N U_0}{2 \pi} \left\{
\int_1^\infty dy \left( \sqrt{ \frac{y^4 - \rho^4}{16 y^4 - 15 + 
2 \rho^4 +    \rho^8}  } - 1 \right) -1 +\rho +
 \frac{1}{4} \sqrt{1 - \rho^4}\right\}.
\ee
We therefore, reach the positive energy condition before we reach
$L_{max}$.

It should be emphasis that the configuration where the D5-brane is at
the horizon is static only from the field theory point of view. 
Namely, unlike the  configuration where the D5-brane is above the
horizon which is static because the net
force at the vertex (including the gravitational one) is zero,
here the net force on the vertex is positive.
Therefore, the D5-brane falls into the black hole.
However, from the point of view of an observer located at the boundary
(a field theory observer)
it takes the D5-brane an infinite amount of time to cross the horizon
hence the configuration is static.
We should also note that since the D5-brane is a freely falling object
it will not be burned by the uge Hawking temperature at the horizon.

To summarize
the behavior of the total energy as a function of the size $L$ is
similar to the case of the quark-anti-quark pair. For small size we
find a Coulomb like behavior but at a certain critical size $L_c$ the
energy becomes zero and we find that the baryon should decay into a
configuration of $N$ quarks with vanishing interaction.

{\it Baryons in non-conformal field theories with sixteen supercharges}

As was explained in the introduction one can generalize the 
supergravity construction of the baryons to the non-superconformal
theories living on a collection of $N$ Dp-branes (for $p\neq 3$)
using the relevant supergravity solution \cite{imsy}.
The relation between the size of the baryons and the energy is
basically $N$ times the quark anti-quark potential found in
\cite{juan1,bisy1,rty,bisy2} (for $p\neq 5$),
\be
E_{bar}\sim -N\left( \frac{g_{YM}^2N}{L^2}\right) ^{1/(5-p)}.
\ee

\section{Baryons in non-SUSY theories}

We discuss YM in three dimensions (the generalization to the four
dimensional case is straight forward).
The results which we obtain  are  expected
from the field theory point of view and were anticipated in \cite{og}.
The supergravity solution associated with pure YM in three dimensions is
given by the near-extremal D3-branes solution in the decoupling limit
\be
ds^2 = \frac{U^2}{R^2} \left[ -(1 - \frac{U_T^4}{U^4}) dt^2 +
  \sum_{i=1}^3 dx_i^2 \right] + \frac{R^2}{U^2 (1 -
  \frac{U_T^4}{U^4})} dU^2 + R^2 d\Omega_5^2 ~. 
\ee
To obtain three dimensional theory we need to go to the IR limit and
to consider distances (along $x_1, x_2, x_3$) which are much larger
then $1/T$.
At the region where we can trust the supergravity solution, $R^2\geq
1$, the theory is not quite three dimensional because the QCD string
can probe the compactified direction \cite{witten2,bisy2}.
Nevertheless this theory possesses some properties of YM in three
dimensions \cite{witten2,bisy2,li,oz}.

\begin{figure}
\begin{center}
 \resizebox{8cm}{!}{
   \includegraphics{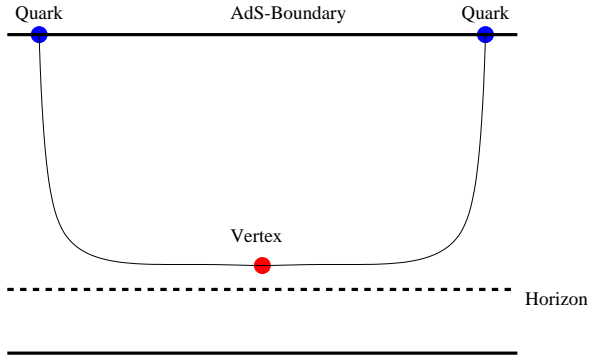}
   }
\end{center}
\caption{The baryon in non supersymmetric theories}
\end{figure}

The surface term gives 
\be\label{suoor}
\frac{1}{4}= \frac{U_0'}{(1-U_T^4/U_0^4)
\sqrt{U^4/R^4+(U_0')^2/ (1-U_T^4/U_0^4)}}.
\ee
To go to the IR limit we need to consider large $L$.
As in the Wilson loop case this means that $U_0\rightarrow U_T$.
At this limit the integrals of $L$ and $E$ are controlled by the region
near $U_0$ and their ratio is a constant which determined   
the QCD string tension.
At first sight it seems that eq.(\ref{suoor}) will change dramatically
the relation between $E$ and $L$.
However, at the IR  limit ($U_0\rightarrow
U_t$)  eq.(\ref{suoor}) implies that $U_0'$ 
 vanishes so the relation is again, as expected, linear
\be
E=N T_{YM} L, ~~~\mbox{where} ~~~
T_{YM} = \pi R^2 T^2.
\ee 

We should note the same relation holds for non-supersymmetric 
YM in four dimensions with the string tension found in  \cite{bisy2}.

\section{Baryons with $k<N$ quarks}

Next we would like to study baryons made of  $k$ quarks when
$k<N$.
For example the case $k=N-1$ gives rise to a baryonic configuration in
the anti-fundamenatl representation.
In a confining theory we do not expect to find such a state (it cost
 an infinite
amount of energy to separate $N-k$ quarks  
all the way to infinity leaving behind the k-baryonic system).
Indeed, as we shall see,
in the non-supersymmetic theories this $k<N$ baryon configuration is
excluded.Surprisingly in the ${\cal N}=4$ theory we do find such
stable k-quarks baryon if $5N/8<k\leq N$.
This is unexpected result which we do not really understand from the
field theory point of view.

The way supergravity enables us to construct  
baryons with less quarks is illustrated in figure 4.
In this figure we have
the usual baryonic vertex with $k$ strings stretched out to the
boundary at $U=\infty$ and the rest $N-k$ strings reaching $U=0$.\footnote{Configurations with strings
  ending on $U=0$ were also considered in the context of quark monopole
  potential \cite{min}}.
 
This configuration is stable provided that $\frac{dE}{dL}>0$.
The calculation of the energy of this configuration proceeds in a
similar way to the calculation of the energy of $k=N$ baryonic system
carries in section 2.
the surface term gives now the following relation

\begin{figure}
\begin{center}
 \resizebox{8cm}{!}{
   \includegraphics{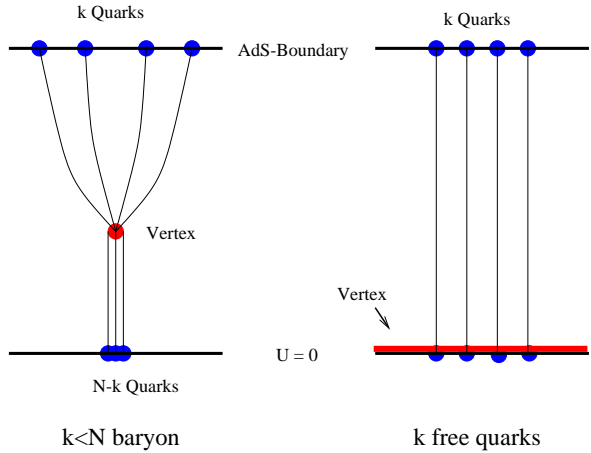}
   }
\end{center}
\caption{The $k < N$ ``baryon'' vs. $k$ free quarks.
Since the longitudinal metric vanishes at $U=0$ the the surface $U=0$
is in fact a point and hence the vertex is smeared along this
``surface'' $U=0$. As a result the string can move freely at the boundary.}
\end{figure}

\be\label{slope1}
\frac{U_x}{\sqrt{U_x^2+U^4/R^4}}= A ,~~~ \mbox{where}~~~
A=\frac{5N-4k}{4k}. 
\ee
For $k=N$ we get $A=1/4$ and for $k<N$ we have $A>1/4$.
It follows from (\ref{slope1}) that $A\leq 1$.
The upper bound , $A=1$ corresponds to $U_x\rightarrow\infty$ and $k=5N/8$.
since the strings are radial the baryon size vanishes.

The energy of the k-quarks baryon is
\be\label{ee}
E_k=\frac{U_0}{2\pi}\left[ (N-k) +N/4+k\left(
\int_1^{\infty} dy ( \frac{y^2}{\sqrt{y^4-(1-A^2)}}-1) -1\right)
\right] .
\ee
Where we have made the same kind of subtraction as in the $k=N$
configuration i.e. we have substracted the energy of $k$ quarks as
depicted in fig.4b.
For $A=1$ ($k=5N/8$) the energy vanishes which implies that the
location of the D5-brane is a moduli of the system.
For $A<1$   the energy is $b U_0$ with some negative $b$ and 
 $U_0$ is determined, as usual,  in terms of $L$.

The fact that no k-quarks baryons exist once $k<5N/8$ can be deduced
by considering the surface relation at the D5-brane.
It is easy to see that in this case not all the $N-k$ strings can go
radially directed towards $U=0$.
Instead they should come out of the vertex with some finite slope and
will therefore, never reach $U=0$.
Instead they will eventually end on the $U=\infty$ boundary leaving us
with more quarks on this boundary.

We would like to end with a short remark on the non-supersymmetic case.
As we remarked at the begging of this section, in a confining field
theory we do not expect to find such states.
This expectation seems to be supported by the AdS supergravity
approach.
The energy of a radial string is 
\be
E=\frac{1}{2\pi}\int_{U_0}^{U_1} dU\sqrt{G_{xx}G_{uu}}\sim \log(U_0-U_T).
\ee
Therefore, the energy of a string stretched between the D5-brane and
the horizon is infinite and hence even the case $k=N-1$  cost an
infinite amount of energy.
Thus the baryonic configuration with $k=N$ is the only stable 
baryonic configuration with finite energy in
 agreement with field theory results.

{\bf Acknowledgement}
We would like to thank O. Aharony for numerous fruitful discussions.

\end{document}